\documentclass{pasj01}
\begin{document}
\title{Dark Supernova Remnant}% revealed by a CO-Line hole}

\author{Yoshiaki \textsc{Sofue}$^{1}$ 
}  
\altaffiltext{1}{Institute of Astronomy, The University of Tokyo, 2-21-1 Mitaka, Tokyo 181-8588, Japan }

\email{sofue@ioa.s.u-tokyo.ac.jp}


\maketitle 

%%%%%%%%%%%%%%%%%%%%%%%%%%%%%%%%%%%%%%%%%%%%%
\def\vlsr{v_{\rm LSR}} \def\Msun{M_\odot} \def\vr{v_{\rm r}}
\def\deg{^\circ}  \def\Vrot{V_{\rm rot}}  
\def\CO{$^{12}$CO$(J=1-0)$ } \def\Htwo{H$_2$} 
\def\Xco{X_{\rm CO}}  \def\Ico{I_{\rm CO}}   \def\Tb{T_{\rm B}}
\def\Htwo{H$_2$ }  \def\Msun{M_\odot} \def\deg{^\circ}  \def\deg{^\circ} 
 \def\co{$^{12}$CO  }  \def\coth{$^{13}$CO  } \def\coei{C$^{18}$O }   \def\cotw{$^{12}$CO ($J$=1-0) } 
 \def\kms{km s$^{-1}$} \def\Kkms{K \kms }  \def\Ico{I_{\rm CO}} 
\def\mH{m_{\rm H}}  \def\Htwo{H$_2$ }      \def\Tb{T_{\rm B}} 
\def\Ico{I_{\rm CO}} 
\def\htwo{H$_2$} 
\def\vlsr{v_{\rm lsr}}  \def\Xco{X_{\rm CO}} 
\def\mum{$\mu$m}
\def\red{}

%%%%%%%%%%%%%%%%%%%%%%%%%%%%%%%%%%%%%%%%%%%%%%%%%%%%%%%%%        
   
\begin{abstract}   
 An almost perfect round hole of CO-line emission with a diameter of 3.7 pc was found in a molecular cloud (MC) centered on G35.75-0.25 ($ l = 35\deg.75, b = -0\deg.25 $) at radial velocity of 28 \kms.
The hole is quiet in radio continuum emission, unlike the usual supernova remnants (SNR), and the molecular edge is only weakly visible in 8 and 24 \mum\ dust emissions.
\red{The hole may be either a fully evolved molecular bubble around a young stellar object (YSO), or a relic of a radio-quiet SNR that has already stopped expansion after rapid evolution in the dense MC as a buried SNR.
Because G35.75 exhibits quite different properties from YSO-driven bubbles of the same size, we prefer the latter interpretation.}
Existence of such a "dark" SNR would affect the estimation of the supernova rate, and therefore the star formation history in the Galaxy. 
\end{abstract}    
 
%%%%%%%%%%%%%%%%%%%%%%%%%%%%%%%%%%%%%%%%%%%%%%%%%%%%%%%%
\section{Introduction}
%http://cda.cfa.harvard.edu/cscweb/index.do

Supernovae (SN) exploded in the Galactic disc with the mean gas density of $\sim 1$ H cm$^{-3}$ expand to form supernova remnants (SNR), which emit thermal and non-thermal radiations due to the strong shock wave during their life time of $\sim 10^4$ y, and they finally merge into the interstellar medium (ISM) at radii exceeding a few tens of pc after $\sim 10^5$ y \citep{chevalier1977,raymond1984,weiler+1988,inoue+2012}.
Such SNR are usually observed as extended objects, often having shell structures, by radio, X-ray, and optical measurements, and about 300 SNRs are currently catalogued 
\citep{green2009}.

Because of the short life time of high-mass stars responsible for SN explosions, SNRs are supposed to be located near to their birth places and adjacent MCs \citep{wheeler+1980,cox1999,seta+2004}.
SNe exploded in MCs evolve as buried SNRs quite differently from those exploded in the inter cloud space, diminishing in a short life time with a small radius \citep{shull1980,wheeler+1980,lucas+2020}.
However, no observational evidence has been obtained yet of such 'buried' SNRs, mainly because of their compactness and short life time.
This implies that the current estimation of the SN rate from the catalogued SNRs in the Galaxy would have been significantly under-estimated.

In spite of the short life time of the buried SNR, the molecular cavity left after the rapid SNR evolution can survive for much longer time on the order of $t\sim r/\sigma_v\sim $ a few pc / a few \kms $\sim 10^6$ y, which would be detectable as a hole in the MC by molecular line observations.
Here, $r$ is the radius of the shell, and $\sigma_v$ is turbulent velocity in the cloud.

In this Letter, we report the discovery of an almost perfect round-shaped hole in a medium sized MC during the search for the sites of SNR-MC interaction \citep{sofue+2020snr},
using the FUGIN (Four-receiver unbiased galactic imaging survey with the Nobeyama 45-m telescope) survey data in the $^{12}$CO, $^{13}$CO, and C$^{18}$O ($J=1-0$) line emissions \citep{ume+2017}. 
The full beam width at half maximum of the telescope was $15''$, the velocity resolution was 1.2 \kms. The final 3D FITS cube data had an  effective beam size of 20$''$, rms noise level of $\sim 1$ K, and a pixel size of ($\Delta l, \Delta b, \Delta \vlsr) = (8''.5, 8''.5, 0.65$ \kms).

 	\begin{figure*}     
 	\begin{center}
\includegraphics[width=13cm]{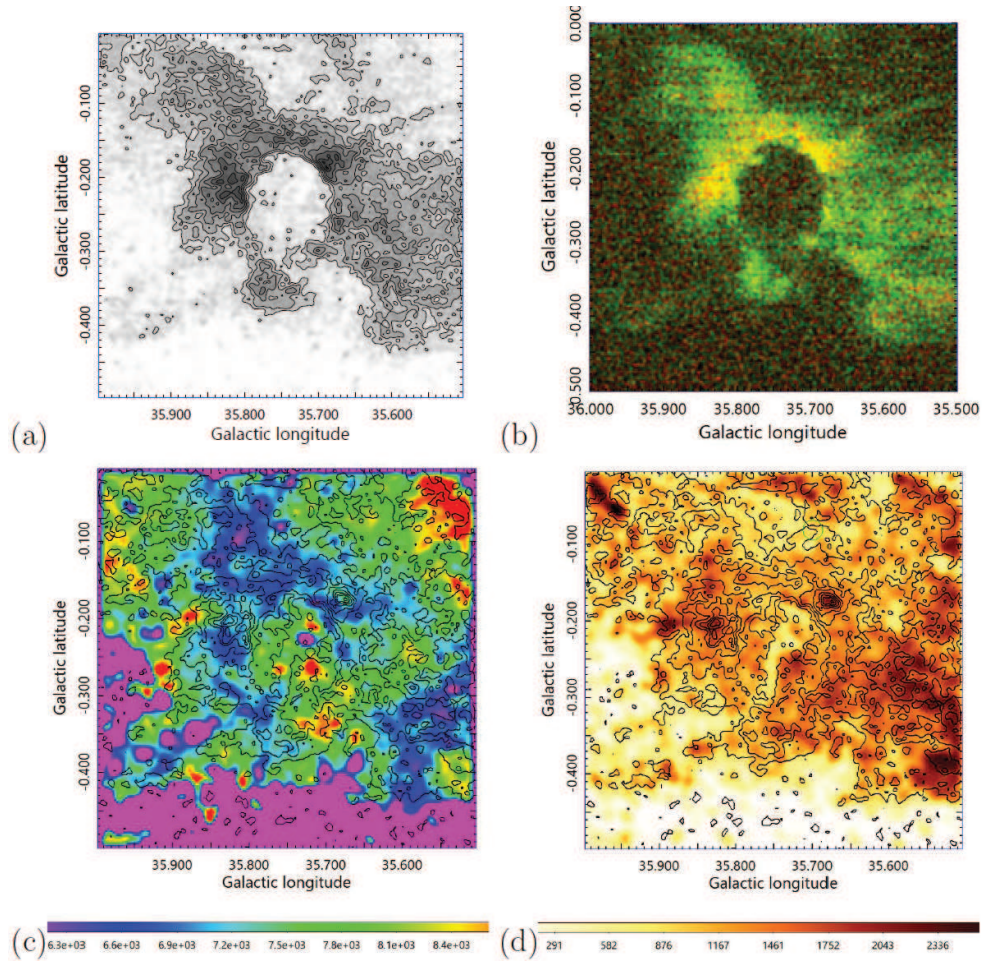}     
\end{center}
\caption{\red{(a) Integrated intensity of the \co line from 24 to 31 \kms around the peculiar CO hole G35.75-0.25+28 \kms. Contours start from 10 with increment 5 \Kkms.
(b) Composite color map of the brightness temperature $\Tb$ of the \co line at 27.775 \kms in green and \coth in red.
(c) Moment 1 and (d) 2 maps showing velocity field (color code in m s$^{-1}$) and velocity dispersion superposed by intensity contours.  }
}
\label{map}    
\end{figure*}     
   
 	\begin{figure}     
 	\begin{center}  
\includegraphics[width=8cm]{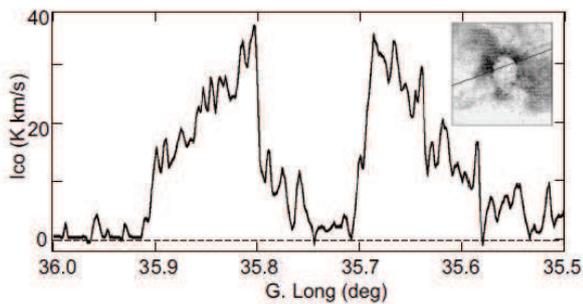}
\end{center}
\caption{Cross section of the \co $\Tb$ across the line as inserted, showing that the hole's center is at the same level as the surrounding region. Such low brightness is explained only if the hole is open along the line of sight. }
\label{cross}   
\end{figure}

 	\begin{figure*} 
\begin{center}            
\includegraphics[width=15cm]{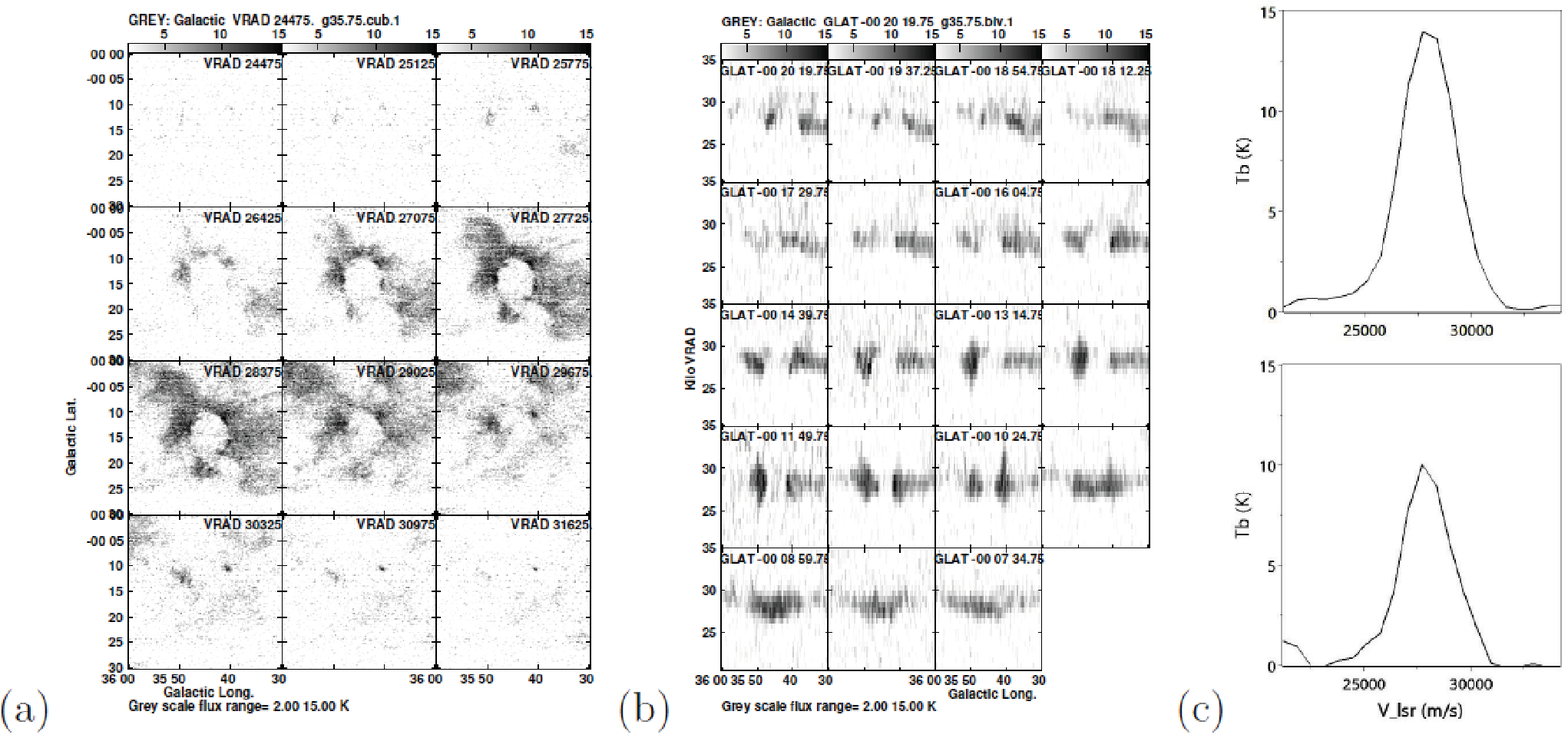}   
\end{center} 
\caption{(a) Channel maps and (b) longitude-velocity diagrams of \co line brightness across various latitudes. (c) \co line profiles at the intensity maximum at the north-west edge (G35.69-0.18) and at a typical position in the north-east CO envelope (G35.80-0.10). } 
\label{cub} 
	\end{figure*}   
	
%%%%%%%%%%%%%%%%%%%%%%%%%%%%%%%%%%%%%%%
\section{Peculiar CO hole G35.75-0.25} 

Figure \ref{map} shows the distribution of integrated \co intensity around the center velocity of 28 \kms centered on G35.75-0.25 ($ l = 35\deg.75, b = -0\deg.25 $), which shows a round-shaped hole of CO emission in a molecular cloud.
Figure \ref{cross} shows the cross section of intensity across the hole.
The figures show an almost perfect, round-shaped hole penetrating the molecular cloud with a steep edge revealed by the sharp intensity cut off. 
The surrounding molecular gas density is highest near the hole's edge, and decreases outward.
Accordingly, the \coth emission in red has the maximum at the highest density regions with extended \co envelope in green.
The cross section indicates that the hole's edge is not compressed against the surrounding cloud, but is sharply cut from a supposed increasing density profile toward the center.
This suggests that the hole was created in the cloud by dissociating the molecules, but not by sweeping the interior gas to the edge.

From the radial velocity at 28 \kms of the CO emission, the kinematic distance is estimated to be $1.8\pm 0.3$ kpc for near-side solution, and $11.1\pm 0.3$ for far solution using the newest rotation curve of the Milky Way \citep{sofue2020}.
Because the molecular cloud surrounding the hole is coincident in position with a dark nebula in the near infrared images (GLIMPSE, http://www.spitzer.caltech.edu/glimpse360/aladin), %\citep{glimpse}, 
we may choose the near distance,  $d=1.8$ kpc. 
The angular diameter of the CO hole is $7'$, which yields a linear diameter of 3.7 pc.

The column density of the molecular gas near the intensity maximum is $N_{\rm H_2}\sim 7\times 10^{21} ~ {\rm H_2 ~ cm^{-2}}$ for a conversion factor of $2\times 10^{20}$ H$_2$ (\Kkms)$^{-1}$ 
\red{for individual molecular clouds \citep{sofue+2020xco}}. 
This yields a volume density of $n_{\rm H_2} \sim 2\times 10^3 ~{\rm H_2 ~ cm^{-3}}$, and total molecular mass of the cloud in the mapped area is estimated to be $\sim 2.4\times 10^3 \Msun$. 
 
\begin{figure*}     
 	\begin{center}   
\includegraphics[width=15cm]{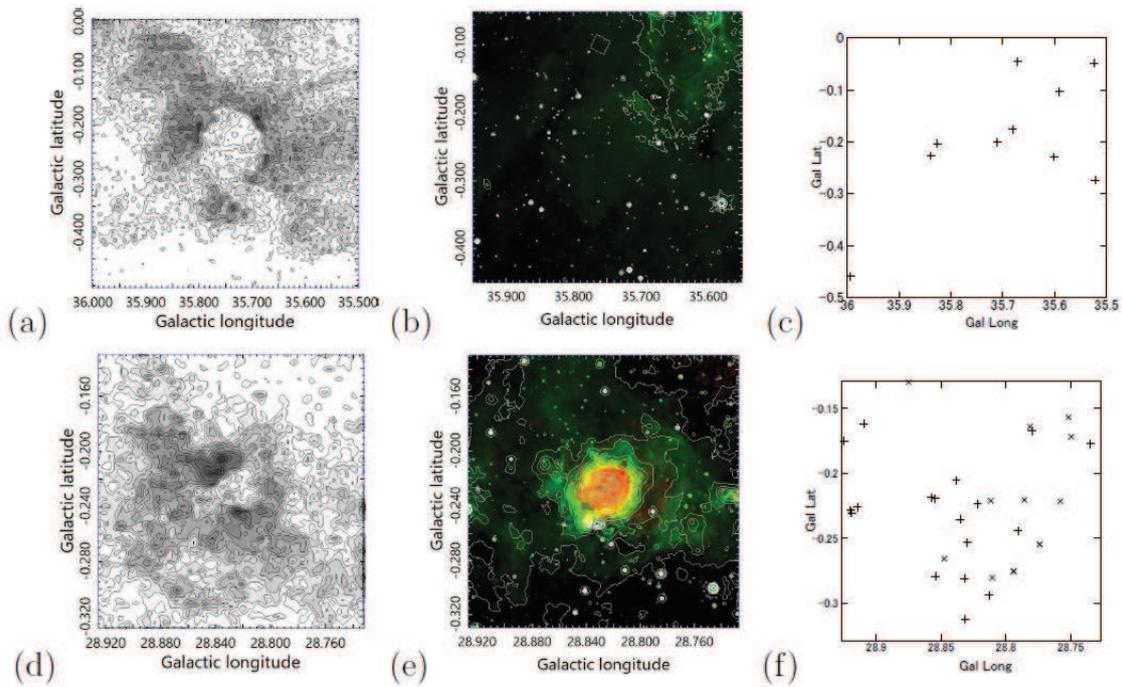}   
\end{center}
\caption{\red{Comparison with a Spitzer bubble N49 of about the same size. 
(a) \co channel map of G35.75 at 28.8 \kms by contours (start 5 K, increment 2.5 K) 
(b) Color-coded map of 20-cm radio in red (1 to 3 mJy/beam), 8 \mum\ in green (40 to 200 mJy/str), and 24 \mum\ by white contours (1 to 1000 mJy/str by 0.1 dex log. increment). 
(c) Distribution of YSOs from SIMBAD marked by crosses (AGAL, 2MASS) (http://simbad.u-strasbg.fr/simbad/).
(d),(e),(f) Same, but for a Spitzer bubble N49 at 85.2 \kms with the same intensity scales\citep{deharveng+2010} with CO at 86.2 \kms, whose linear size is $\sim 4$ pc, about same as G35.75. 
Note the quiet G35.75 in radio and infrared, and much higher YSO density in N49. }
}
\label{yso}    
\end{figure*}

\red{Inspection of the archival data indicates that the hole is not associated with radio continuum emission \citep{chur+2009,stil+2006}.
It is also faint in far infrared emissions, although the north-western edge is slightly bright in 8 and 24 \mum\ (MAGPIS, https://third.ucllnl.org/gps/index.html).
In figure \ref{yso} we present a composite map of the 20-cm radio continuum, 8 \mum, and 24 \mum\ emissions.}

\begin{figure}     
 	\begin{center}  
 \includegraphics[width=8cm]{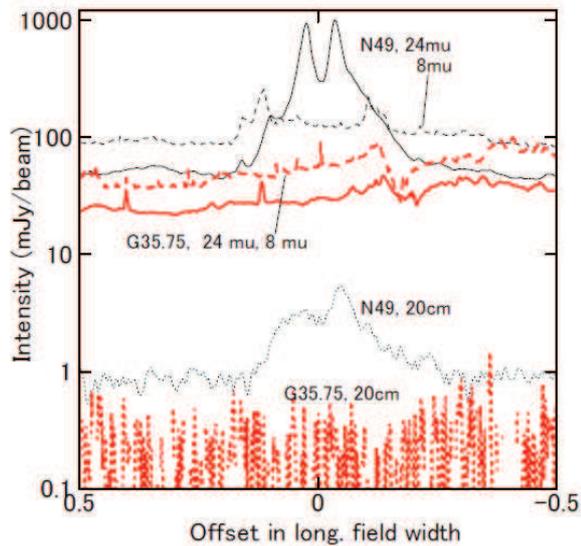} 
\end{center}
\caption{\red{ Cross sections of the intensities across the centers of G35.75 and N49 from figure \ref{yso} relative to the field width in longitude.
Note one to two orders of magnitude lower intensities of G35.75 in radio and infrared, in spite of comparable \co brightness. }
}
\label{compintensity}    
\end{figure}

\red{In order to see how the object is different from a Spitzer bubble, we compare the radio and infrared maps with those of an isolated Spitzer bubble, N49, of $0\deg.05$ in diameter, or 4 pc at a distance of 5.5 kpc \citep{deharveng+2010}, which is also embedded in a similar sized molecular cavity at 86.2 \kms.
It is stressed that the projected number density of YSO around G35.75 is an order of magnitude smaller than that around N49, and that only one YSO inside the hole is significantly off-set from the center.
Figure \ref{compintensity} shows cross sections of the intensities across the CO hole centers, demonstrating that G35.75 is one to two orders of magnitudes fainter in infrared and radio.
We also mark the positions of YSO as obtained from SIMBAD (AGAL, 2MASS: http://simbad.u-strasbg.fr/simbad/).}

\section{Dark SNR?}
 
\red{From these facts, we may consider two possibilities about the origin of G35.75-0.25.
One idea is that it is a fully evolved Spitzer bubble possibly excited by the YSO seen inside the bubble, which is, however, significantly off set from the center.
The other is that it is a relic of a buried SNR in the MC created by SN explosion in the dense molecular gas, where the YSO is located by chance on the line of sight.}

\red{Unless the YSO's distance is determined to coincide with the cavity's, 1.8 kpc, within a few pc accuracy, we may consider it to be there by chance on the sky.
Also, from the well cut inner edge of the bubble as well as for its particularly round shape, we prefer the latter possibility.
So, here, we will try to explain the CO hole as due to a fully evolved buried SNR. }

\red{Although in its pre-explosion phase, the progenitor star would have produced a cavity  by stellar wind, the SN shock front reaches the cavity's boundary in a sufficiently shorter time compared with the entire evolutionary time scale \citep{inoue+2012,lucas+2020}.
Since most of the released energy by core-collapse SN explosion, $\sim 10^{51}$ erg, is exhausted by the initial infrared flash at the shock front against the molecular gas within $\sim 10^2$ years \citep{shull1980}, the kinetic energy given to the molecular gas may be assumed to be on the order of $E_0\sim 10^{50}$ erg.
} 

The evolution of a SNR in its evolved phase can be approximated by an adiabatic shock wave.
Then, the expansion velocity $v$ and radius $r$ of the shock front is approximately related to the input energy $E_0$ and gas density $\rho_0$ as 
$E_0 \sim \frac{1}{2} \left(\frac{4\pi}{3}r^3\rho_0 \right)v^2$.
Remembering $v=dr/dt$, we have a differential equation for $r$ as a function of time $t$, and the solution yields 
$ v\sim 0.7 (E_0/\rho_0)^{1/2} r^{-3/2},$
$ v\sim  1.2 (E_0/\rho_0)^{1/5}t^{-3/5}$, and
$ r\sim 0.5 (E_0/\rho_0)^{1/5}t^{2/5}$. 
This leads to $v\sim 20$ \kms at $r\sim 2$ pc and $t\sim 5\times 10^5$ y for $n_{\rm H_2}=10^4$ cm$^{-3}$ ($\rho_0=\mu \times 10^4$ H cm$^{-3}$, with $\mu=2.8$ ).

\begin{figure}     
 	\begin{center}   
\includegraphics[width=8cm]{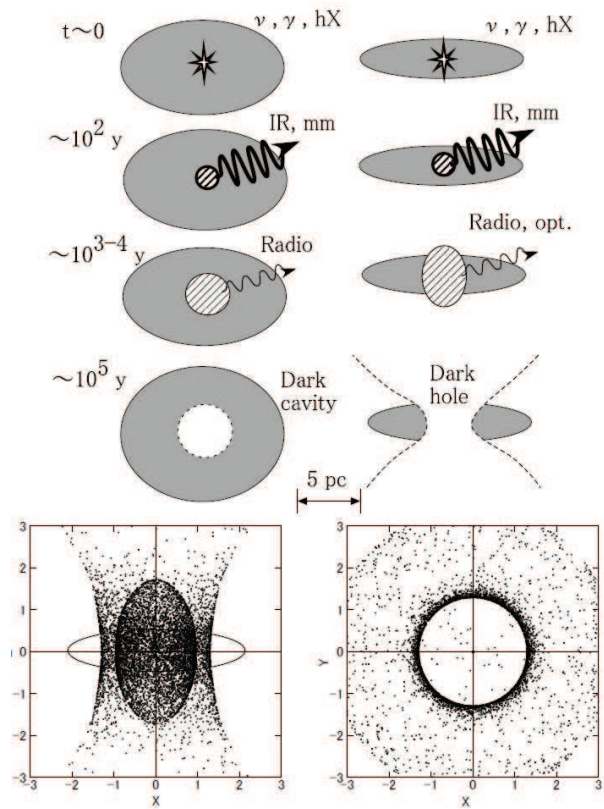} 
\end{center}
\caption{\red{(Top) Schematic evolution of a buried SNR in an extended molecular cloud (left panel) and in a flat cloud finally broken in the polar direction (right panel).
(Bottom) 3D shock front in a flat spheroidal cloud by explosion at the center. Left and right panels show side and top views, respectively. Scale is arbitrary, and see the estimation in the text for the real scale and time.
}}
\label{model}   
\end{figure}

If the cloud is large enough compared to the shell size, the SNR dies, being kept buried in the cloud, leaving a dark cavity, as illustrated in figure \ref{model}.
However, if the cloud is flattened, the shock front is elongated in the polar direction, and finally open to the inter-cloud space. 
Accordingly, the internal pressure escapes to the polar direction, and  the expansion in the equatorial plane will be decelerated more rapidly than in the spherical expansion.

Such a hole seen from the pole will appear as a round cavity having brightness as low as the surrounding background, as indeed observed (figure \ref{cross}).
Since the SNR's energy is almost exhausted at this stage, no particular kinematic disturbance to increase the velocity dispersion is observed in the cloud (figures \ref{map}(d), \ref{cub}(c)). 
\red{Even if the gas is expanding at the velocity as estimated above, since the motion is in the plane of the sky perpendicular to the line of sight, the velocity width at the edge will be not much increased compared to the ambient velocity dispersion, as indeed observed (figure \ref{cub}(c)).
This makes a contrast to the high velocity wing observed in a molecular cloud interacting with the bright SNR such as W44, which is expanding at $\sim 100$ \kms into the inter-cloud space with much lower density \citep{seta+2004}.}

3D evolution of the shock front can be simulated by applying the radial adiabatic shock method \citep{sakashita1971,sofue2019}, as shown in figure \ref{model}, where a point explosion takes place at the center of a cloud having a Gaussian density profiles of radius 2 and thickness 0.5 in arbitrary units.
As the time elapses, the shock front expands into the polar direction, and breaks the cloud to create a hole, mimicking the observed molecular hole having a clear-cut inner edge (figure \ref{cross}).
 
%%%%%%%%%%%%%%%%%%%%%%%%%%%%%%%%%%%%%%%%%%%%%%
\section{Summary}
 
 A round-shaped hole of CO-line emission with a diameter of 3.7 pc was found in a molecular cloud centered on G35.75-0.25 at $\vlsr=28$ \kms, whose center is almost empty in radio and infrared emissions. 
The hole is interpreted as a relic of a fully evolved SNR that has already stopped expanding after its rapid evolution in the dense molecular gas as a buried SNR. 
Detection of such a "dark" SNR gives evidence for the existence of buried SNRs in the Galactic plane, probably every where.
This would imply that the supernova rate in the Milky Way based on optical, radio and/or X-ray bright SNRs would have been significantly underestimated. 
  
\vskip 5mm
\noindent {\bf Acknowledgements}:
The CO data were taken from the FUGIN CO survey obtained with the Nobeyama 45-m telescope (http://nro-fugin.github.io).  Data analysis was carried out at the Astronomy Data Center of the National Astronomical Observatory of Japan.

\end{document}